\documentclass{sadhana}

\usepackage{graphicx,amsmath,bm}
\usepackage{supertabular}
\usepackage{color,soul}

\begin{document}

\title{Enhancing the Accuracy of Transition Models for Gas Turbine
Applications Through Data-Driven Approaches}


\author{Harshal D. Akolekar\textsuperscript{1,}\textsuperscript{2}\textsuperscript{*}}
\affilOne{\textsuperscript{1} Department of Mechanical Engineering, Indian Institute of Technology Jodhpur, Rajasthan, India, 342030\\}
\affilTwo{\textsuperscript{2} Department of Mechanical Engineering, University of Melbourne, Victoria, Australia, 3010}


\twocolumn[{

\maketitle

\begin{abstract}
Separated flow transition is a very popular phenomenon in gas turbines, especially low-pressure turbines (LPT). Low-fidelity simulations are often used for gas turbine design. However, they are unable to predict separated flow transition accurately. To improve the separated flow transition prediction for LPTs, the empirical relations that are derived for transition prediction need to be significantly modified. To achieve this, machine learning approaches are used to investigate a large number of functional forms using computational fluid dynamics-driven gene expression programming. These functional forms are investigated using a multi-expression multi-objective algorithm in terms of separation onset, transition onset, separation bubble length, wall shear stress, and pressure coefficient. The models generated after 177 generations show significant improvements over the baseline result in terms of the above parameters. All of the models developed improve the wall shear stress prediction by 40-70\% over the baseline laminar kinetic energy model. This method has immense potential to improve boundary layer transition prediction for gas turbine applications across several geometries and operating conditions. 
\end{abstract}



}]


\setcounter{page}{1}
\corres
\volnum{- }
\issuenum{ -}
\monthyear{ -}
\pgfirst{- }
\pglast{ -}
\doinum{- }
\articleType{}


\markboth{Harshal Akolekar}{Enhancing Accuracy of Transition Models for GT Applications Through Data-Driven Approaches}

\section{Introduction}
As gas turbine design has reached an advanced stage of development, innovative approaches in the design phase are imperative to achieve further gains in efficiency. A critical component impacting the overall efficiency of gas turbines is the Low-Pressure Turbine (LPT). Michelassi et al. \cite{Michelassi2016} have reported that even a 1\% increase in LPT efficiency can lead to a noteworthy 0.6\%-0.8\% reduction in specific fuel consumption, translating into substantial fuel savings and emission reductions. As an aircraft ascends and enters cruise mode, the Reynolds number associated with the LPT decreases dramatically, shifting from approximately $5 \times 10^5$ to $0.5\times 10^5$ \cite{Mayle1991}. Over this wide range of Reynolds numbers, there is a significant alteration in flow dynamics, making LPT design a challenging endeavor. During take-off, the blade boundary layers are primarily turbulent, while during cruise, extensive laminar regions emerge, and the transition of boundary layers plays a pivotal role in determining engine performance. The state of the boundary layer significantly influences the overall efficiency of an LPT \cite{Hodson2005}.

The transition process occurs in three primary modes: natural, bypass, and separated flow. The latter two are often significantly affected by unsteadiness, particularly in the form of incoming wakes, which is highly relevant to LPTs \cite{Hodson2005}. In LPTs, separated-flow transition is one of the most common modes of transition. When a laminar boundary layer separates, transition may manifest within the free-shear-layer-like flow in close proximity to the blade surface \cite{Pacciani2011}. 
The flow can reattach and form a laminar separation bubble, develop into a turbulent reattachment bubble, or remain detached, leading to an open separation.
The flow may reattach, forming either a laminar separation bubble or a turbulent reattachment bubble, or it may stay detached, leading to an open separation. In LPTs, this transition in separated flow usually occurs in areas with adverse pressure gradients near the leading edge or after peak suction. Predicting this transition accurately is vital for designing high-lift blades, which are especially prone to separation due to their increased flow turning angles. It directly influences the size and behavior of separation bubbles at lower Reynolds numbers, making it a crucial consideration in design.

With the continual increase in computational power, the feasibility of high-fidelity simulations capable of accurately predicting separated-flow transition and wake mixing has grown significantly \cite{Sandberg2022,Sandberg2019}. However, these simulations remain impractical as industrial design tools due to their high computational costs. Consequently, the design of LPTs continues to rely heavily on Reynolds-Averaged Navier-Stokes (RANS) calculations, which offer a cost-effective means of modeling transitional flows for engineering applications.

According to Denton \cite{Denton1993}, many computational fluid dynamics (CFD) calculations struggle to precisely predict the onset point of transition. To address this challenge, a practical yet not universally applicable approach involves prescribing the transition location as a fixed point on the blade surface, typically somewhere downstream from the point of peak suction. In recent years, efforts have been made to enhance the accuracy of transition prediction within RANS calculations, particularly in relation to intermittency ($\gamma$).
and transition momentum thickness ($Re_{\theta}$)\cite{Menter2006}.
One notable model, the $\gamma$-$Re_{\theta}$ model, incorporates two transport equations for $\gamma$ and $Re_{\theta}$, alongside the treatment of turbulent kinetic energy (TKE) and specific dissipation rate ($\omega$). Menter et al. \cite{menter2015one} also released the one-equation $\gamma$ model which only solves one additional transport equation and removes all non-local modeling. Additionally, a different class of models leverages the concept of laminar kinetic energy (LKE)\cite {Mayle1997,Pacciani2011}. These models involve solving transport equations for LKE, TKE, and $\omega$. There is also the $k\overline{v_2}\omega$ model \cite{Lopez2016a}, which addresses transport equations for TKE, $\overline{v_2}$ (representing wall-normal velocity fluctuations initiating transition), and $\omega$.

However, it is important to note that existing RANS-based transition models encounter difficulties in accurately predicting either the onset of transition, length of separation bubbles on the suction side, or turbulence ignition peak amplitude of LPTs. Consequently, there is a pressing need to explore innovative models and methods to enhance the precision of transition prediction within RANS calculations. Data-driven approaches are a very novel approach to improve RANS modeling \cite{Duraisamy2018,Brunton2020}. 
In this study, using CFD-driven \cite{Zhao2020} multi-expression multi-objective (MEMO) \cite{Waschkowski2021} gene expression programming (GEP) for turbulence modeling \cite{Weatheritt2016}, a number of functional forms based on non-dimensional $\Pi$ groups \cite{Akolekar2021,Fang2024} are developed to predict separated-flow transition for LPTs. The accuracy of these novel functional forms is investigated using Pareto analysis and is assessed in terms of boundary layer parameters such as wall shear stress, separation and transition onsets, separation bubble length, and the pressure coefficient plateau in the trailing edge boundary layer.

The paper is structured as follows. Section 2 of this study discusses the methodology in detail which outlines the numerical setup, the baseline LKE model, and the correction terms. Section 3 comprises the results and discussions and how the newly developed model formulations have the potential to improve transition flow prediction. Section 4 concludes the study and gives a futuristic perspective on the field. 

\section{Methodology}
This section describes the baseline transition model, the proposed functional forms, the numerical setup, and the machine learning algorithm.

\subsection{\textbf{Numerical Setup}}
The study utilized the T106A Low-Pressure Turbine (LPT) linear cascade, shown in Fig. 1, as the foundation for model development. The research involved steady, compressible, two-dimensional (2D) RANS simulations with consistent steady inflow conditions. These simulations were conducted using the TRAF solver \cite{arnone1994}, which is written in FORTRAN90. The primary focus was on the midspan section of the cascade, as it significantly contributes (30\%-40\% \cite{Denton1993}) to the overall losses in LPTs.
The primary focus of this study was the midspan section of the cascade since it plays a substantial role in contributing to the overall losses in LPTs. 
The choice of flow configuration was made to ensure that the isentropic exit conditions, including Reynolds number ($Re_{2is} = 100,000$) and Mach number ($Ma_{2is} = 0.4$), as well as the inflow conditions characterized by turbulence intensity (Tu1 = 4\%) and the inlet flow angle ($\alpha_1 = 46.1^\circ$), direct numerical simulation (DNS) data presented by Michelassi et al. \cite{Michelassi2016}.
The DNS data used in this paper has been well validated
against experiments \cite{Stadtmuller2001}. Additionally, accurate wall shear stress data is not
available from experiments. Hence DNS is used as the reference data.
To maintain consistency, the non-dimensional turbulent inlet length scale ($l_t$ / $C_{ax}$) was set at 7.5$\times$10$^{-3}$ throughout the simulations. The spatial discretization was done with second-order schemes. An O-H type grid was used and extensive grid convergence studies have been conducted here \cite{Akolekar2019thesis}. 
\begin{figure}[htb]
\centering
\includegraphics[scale=0.43]{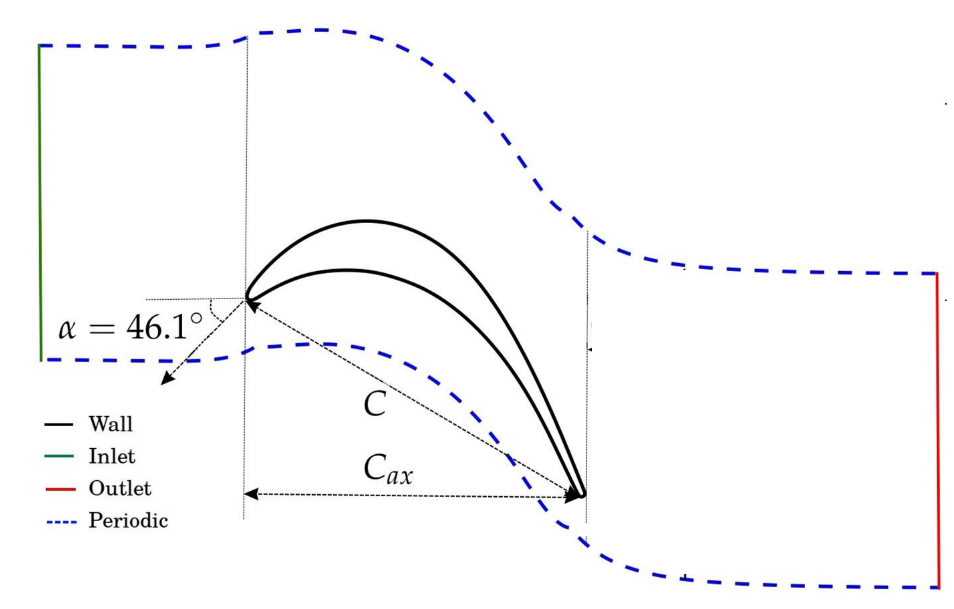}
\caption{The T106A LPT blade domain.}
\label{fig3}
\end{figure}

\begin{figure*}[htb]
\centering
\includegraphics[scale=0.8]{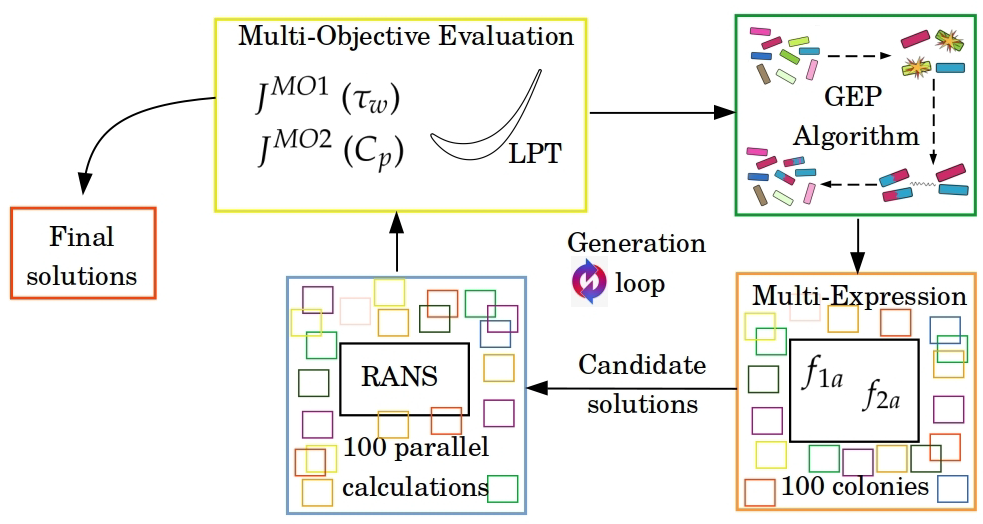}
\caption{The CFD-driven MEMO algorithm for transition model development.}
\label{fig3}
\end{figure*}

\subsection{\textbf{The Laminar Kinetic Energy Model}}
The baseline transition model that is used is the laminar kinetic energy model (LKE - $k_l$), which consists of three transport equations \cite{Pacciani2013}: 
\begin{equation}
\begin{aligned}
&\frac{D k}{D t} =  P_k - \beta^*f_k k\omega + \frac{\partial}{\partial{x_j}}\left[(\nu+\delta_k \nu_T)\frac{\partial{k}}{\partial{x_j}}\right] + R, \\
& \frac{D k_l}{D t} =  P_l - \epsilon_l + \nu \nabla^2 k_l - R, \\
&\frac{D \omega}{D t} =  \alpha \frac{\omega}{k}P_k - \beta^*\omega^2 + \frac{\partial}{\partial{x_j}}\left[(\nu+\delta_{\omega} \nu_T)\frac{\partial{\omega}}{\partial{x_j}}\right].
\label{eq2}
\end{aligned}
\end{equation}
For the $k_l$ transport equation, the unsteady and convection terms are on the left-hand side, and the production $P_l$, dissipation ($\epsilon_l=2\mu{k_l}/{y^2}$), diffusion ($\nu \nabla^2 k_l$) and transfer terms ($R$) are on the right-hand side. $R$ represents the LKE transfer to turbulent kinetic energy (TKE - $k$) and is used to couple the transition model with the $k-\omega$ turbulence model. 
$P_l$ controls the production of LKE and R adjusts the transfer of LKE to TKE.
They are thus pertinent terms in the model and will have a direct effect on the transition prediction.
Hence, we will attempt to use ML to find improved formulations for $P_l$ and R. The production term $P_l$ (see Eq. \ref{eq3}) is formulated with the laminar eddy viscosity $\nu_l$ and the mean shear rate $S$. This is consistent with the latest research \cite{Dick2017} indicating the laminar fluctuations are amplified mainly due to the conventional shear-stress/strain interaction. The original laminar eddy-viscosity ($\nu_l$) is constructed with an estimator of the shear-layer vorticity thickness $\delta_{\Omega}$, $k_l$ and a constant $C_1$:
\begin{equation}
\begin{aligned}
& P_l=\nu_lS^2, \\
&\nu_l = C_1\sqrt{k_l}\delta_\Omega, \\
& \delta_{\Omega}= \textrm{min}\left(\frac{\Omega y^2}{U},2\right). 
\label{eq3}
\end{aligned}
\end{equation}
The transfer term $R$ represents the process of LKE being transferred to the TKE. The original $R$ term is
\begin{equation}
\begin{aligned}
 &  R=C_2 f_{t2} \omega k_l, \\
 & f_{t2}=1-e^{\psi/C_3},\\
 & \psi=\textrm{max}(\underbrace{R_y}_{\textrm{transition sensor}}-\underbrace{C_4}_{\textrm{threshold}},0), \\
 & R_y=\frac{\sqrt{k}y}{\nu}.
\end{aligned}
\label{eq4}
\end{equation}
By the formulation of $\psi$ as depicted in Eq. 3, the commencement of transition occurs when the flow feature sensor - i.e.  Reynolds number based on wall distance ($R_y$), attains the cutoff value denoted as $C_4$, where $C_4 = 10$.  

\subsection{\textbf{Non-Dimensional $\Pi$ groups}}
To improve the transition prediction, the formulation of the $\psi$ and $\nu_l$ terms will be improved. The current formulation for $\psi$ is quite basic for triggering transition onset and $\nu_l$ includes non-local terms and are not ideal for transition model purposes. Even Menter et al. \cite{Menter2006} revised their model to remove non-local formulations \cite{menter2015one}.  
These terms will be modified using seven non-dimensional $\Pi$ groups \cite{Akolekar2021,Akolekar2022} which have been derived from local independent dimensional variables using the Buckingham Pi theorem \cite{white2017}:
\begin{equation}
\begin{aligned}
     & \Pi_1 = \frac{k_l}{\nu\Omega}; \hspace{1mm} \Pi_2  \frac{\Omega y}{U}; \Pi_3 =  \frac{y}{l_t}; \hspace{1mm} \Pi_4 = \frac{\sqrt{k} y}{\nu} \\
    & \Pi_5= \frac{k}{\nu\Omega}; \hspace{1mm} \Pi_6=\frac{\sqrt{k}}{\Omega y}; \hspace{1mm} \Pi_7=\frac{\omega}{\Omega};
    \end{aligned}
\end{equation}

A local formulation for $\nu_l$ and $\psi$ is proposed as
\begin{equation}
    \nu_l  = f_{1a}(\Pi_i)y\sqrt{k_l} ; \hspace{2mm} \psi = max(f_{2a}(\Pi_i),0)
\end{equation}
where $f_{1a}(\Pi_i)$ and $f_{2a}(\Pi_i)$ are functions of the seven non-dimensional $\Pi$ groups. All of these functional forms are local in nature and can be easily incorporated in a RANS framework and have been used as transition sensors of some kind in other words \cite{Dick2017}. $\Pi_1$  is a ratio of time scales $k_l/\nu \omega^2$ and $1/\omega$. $\Pi_5$ is a ratio of time scales $k/\nu \omega^2$ and  $1/\omega$, respectively. $\Pi_2$ and $\Pi_6$ are relevant scales for the shear layer thickness. $\Pi_3$ is a term for the turbulence fluctuations from the boundary layer. $\Pi_4$ is the wall distance Reynolds number and $\Pi_7$ ratio of turbulent fluctuations time scale ($1/\omega$) and laminar fluctuations time scale ($1/\Omega$). 
These functional forms can be found with the aid of the CFD-Driven machine learning technique \cite{Zhao2020}.  

\subsection{\textbf{CFD-Driven Machine Learning Formulation}}
Figure 2 shows the CFD-driven MEMO algorithm \cite{Waschkowski2021}, based on gene expression programming \cite{Weatheritt2016}, which is written in Python. Each population, with the first population being randomly generated, consists of 100 candidate colonies, which consist of two expressions (individuals) for $f_{1a}$ and $f_{2a}$. The functional form of these expressions is fed into the TRAF solver and 100 RANS calculations are performed for each generation. The entire process has been performed for 177 generations, using close to 10,000 core hours. To evaluate the performance of these colonies, two cost functions, relying on the suction side wall shear stress ($\tau_w$) and pressure coefficient ($C_p$), are employed:
\begin{equation}
\begin{aligned}
    & J^{MO1} = \sum_{x/C_{ax}=0.6}^1 = (\tau_w^{DNS} - \tau_w^{RANS})^2_{SS} \\
    & J^{MO2} = \sum_{x/C_{ax}=0.6}^1 = (C_p^{DNS} - C_p^{RANS})^2_{SS}
    \end{aligned}
\end{equation}
The cost functions are defined over the suction side of the blade, extending from the point of peak suction to the trailing edge. This encompassing range includes the separation plateau observed in the pressure coefficient plot, a characteristic often seen in LPTs experiencing transition induced by separation.

The objective is to minimize these two distinct cost functions. This minimization process involves assessing the fitness of different colonies based on their respective cost function values. Subsequently, a new generation of colonies is created through natural selection and genetic modifications. These updated models are then subjected to comprehensive Computational Fluid Dynamics (CFD) calculations to evaluate their performance.
This training process iterates continuously, leading to the evolution of the population of candidate colonies over 177 generations. The overarching goal remains the same: to minimize both cost functions. It is worth noting that due to the presence of two distinct cost functions, there isn't a single definitive optimal solution. Instead, multiple `best' solutions are identified, each offering a balance across the various cost functions. 
Pareto analysis is the ideal method for assessing the progress of solutions in multi-objective optimization. It relies on Pareto domination, where one candidate dominates another if it has lower or equal cost function values, with at least one being strictly lower \cite{konak2006multi}. Through pairwise comparisons across the entire population, candidates receive ranks \textit{(r)}. In the same rank, no solution dominates another, but all solutions in rank r = m1 dominate those in rank r = m2 (where m1 $<$ m2). Solutions in rank r = 1, known as the Pareto front, are not dominated by any other, signifying their status as non-dominated solutions. 
  
\begin{figure}[htb]
\centering
\includegraphics[scale=0.225]{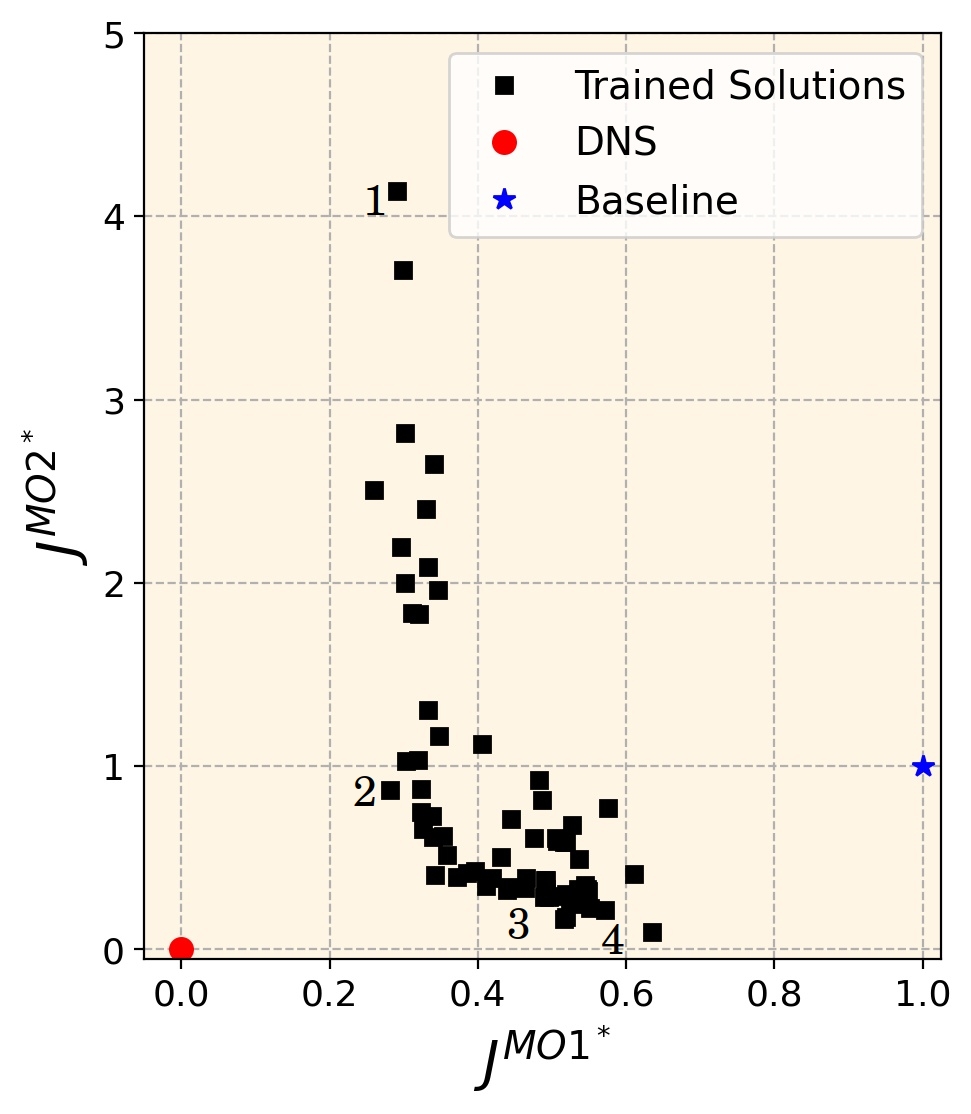}
\caption{Pareto front at the 177$^{th}$ generation.}
\label{pF}
\end{figure}
\section{\textbf{Results \& Discussions}}
\begin{figure}[htb]
\centering
\includegraphics[scale=0.18]{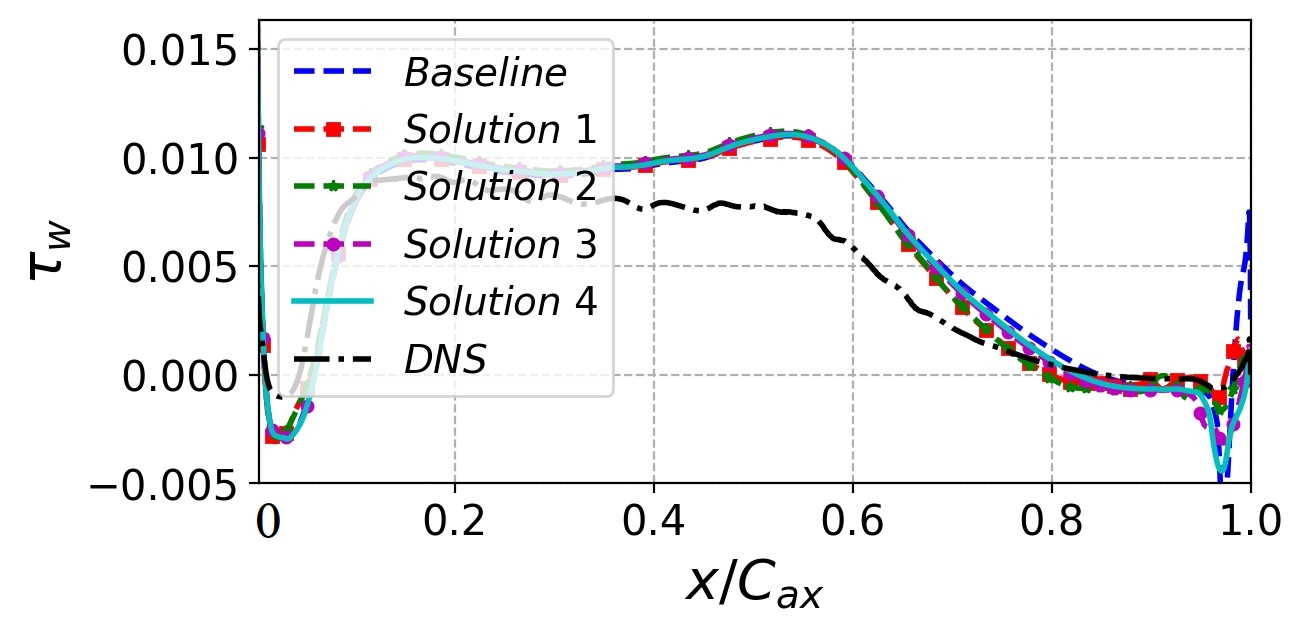}
\includegraphics[scale=0.18]{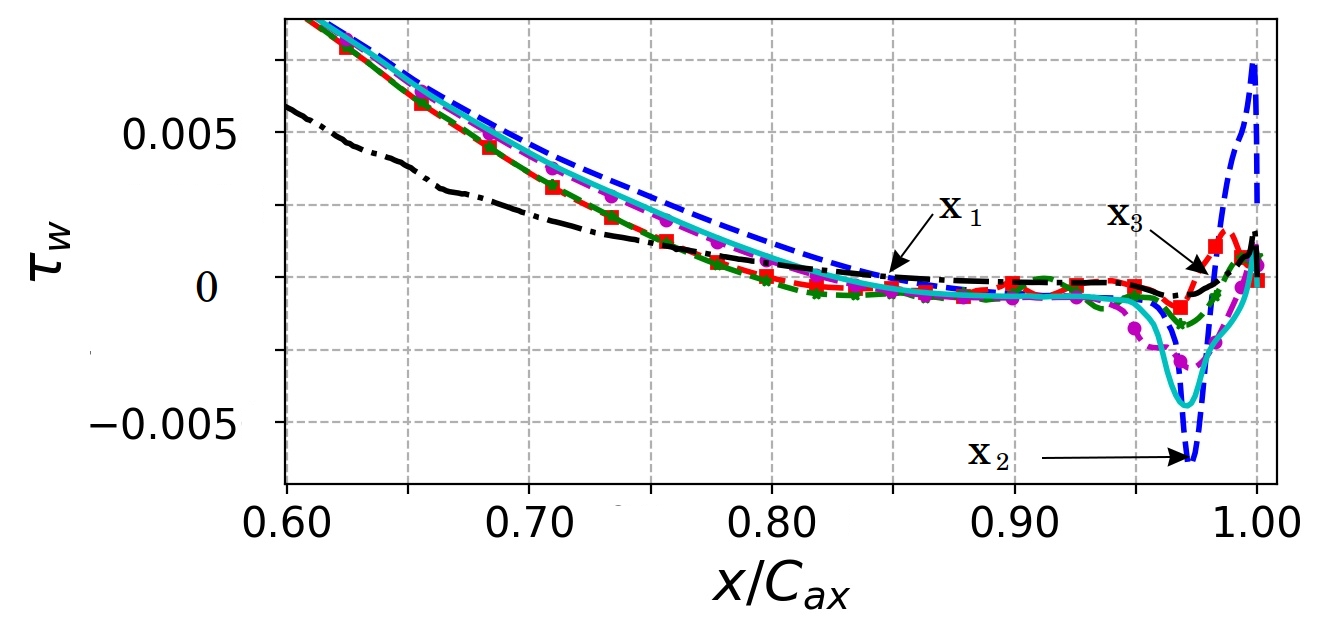}
\caption{Wall shear stress profiles for (a) the suction side, (b) trailing edge region on the suction side.}
\label{wss123}
\end{figure}

With the CFD-driven algorithm, a number of expressions were formulated for $f_{1a}$ and $f_{2a}$. Initially, RANS calculations were conducted with the baseline (B) LKE model which resulted in $J^{M01}_B = 5.24 \times 10 ^{-6}$ and $J^{M02}_B = 3.50 \times 10 ^{-2}$. At $Re_{2is}=100,000$, LPTs have a closed separation bubble (as is evident from Fig. \ref{wss123}). There is also a characteristic separation plateau in the pressure coefficient plot (Fig. 5). The CFD-driven algorithm was then run for 177 generations and a family of solutions was produced that were significantly better than the baseline case. 

Figure \ref{pF} shows the Pareto front with respect to both the cost functions after the 177$^{th}$ generation. In this figure, the cost functions are normalized with the cost function of the baseline LKE model and are represented as $J^{M01*}$ and  $J^{M02*}$. Hence, the point from the baseline solution is shown at (1,1). The DNS solution is the reference solution and hence it is at (0,0). The closer a solution is towards (0,0), it indicates that it is a better solution. Solutions, of course, have to be individually checked for numerical stability and if they are physically correct. In this plot, all solutions along the Pareto front have $J^{M01*} < 1$. All the solutions have $J^{M01*}$ which is 40\%-70\% of $J^{M01}_B$ . However, all solutions do not have $J^{M02*} < 1$. It was shown previously, that to get an improved transition prediction not only should $J^{M01*} < 1$, but also $J^{M02*} < 1$. This is because the accurate prediction of $C_p$ influences the stability of the solution (evident in Fig. 7). Several solutions fulfill these criteria and are analyzed. Four solutions, as shown in Fig. \ref{pF}, have been chosen for further analysis. These solutions lie along various locations across the Pareto front. Solution 1 has the least wall shear stress cost function. However, it performs worse than the baseline model, with a $J^{M02*} > 4$. Solutions 2, 3, and 4 all have $J^{M02*} < 1$, with solution 4's $J^{M02*}$ tending to 0. 

Figure \ref{wss123} shows the wall shear stress variation along the LPT's suction side. The part after peak suction (i.e. $x/C_{ax}=0.6$ - refer to the pressure coefficient plot in Fig. 5), is zoomed in both Figs. 4 and 5. In Fig. 4 $x_1$ is the separation onset location, $x_2$ is the transition onset location, and is also the location of minimum wall shear stress. $x_3$ is the location at which the separation bubble reattaches (as the separation bubble is a closed one at these operating conditions). The length of the separation bubble is $L_s = (x_3 - x_1)$. \begin{figure}[htb]
\centering
\includegraphics[scale=0.49]{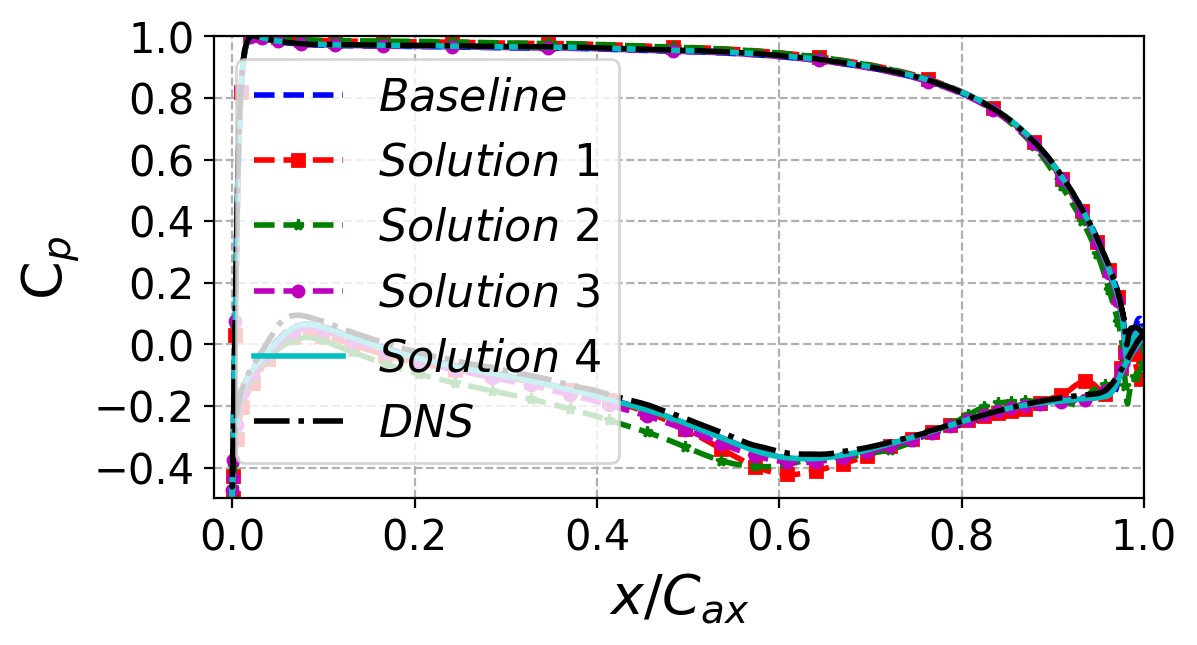}
\includegraphics[scale=0.49]{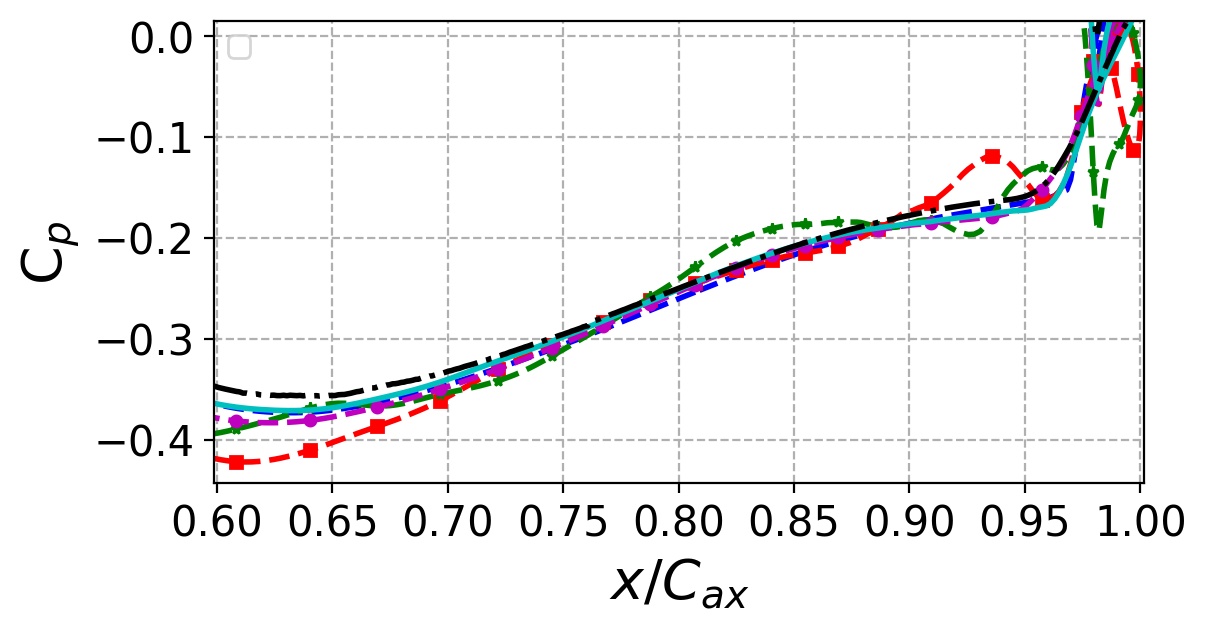}
\caption{Pressure coefficient profiles for (a) the pressure and suction sides, (b) trailing edge region on the suction side.}
\label{cpF}
\end{figure} The DNS result \cite{Michelassi2016}  is shown by the dotted black line.
The baseline result, even though it matches the separation onset quite well, does not capture the transition onset location that well. The wall shear stress profile from $x/C_{ax} > 0.95$ is very far from the DNS result. This drastically affects the flow prediction coming off the blade and the wake mixing prediction downstream of the blade \cite{Akolekar2019a,Akolekar2019,Akolekar2021a,Akolekar2019isabe,Pacciani2021}. Getting the correct wake mixing prediction is crucial for engineers to predict losses in the turbine stage \cite{Praisner2006}. All 4 solutions that have been chosen from the Pareto front improve the prediction of the wall shear stress. However, solution 1 does not improve $J^{MO2*}$ over the baseline solution. Since it does not improve  $J^{MO2*}$, it also leads to a partially unstable solution. Similarly, for solution 2, even though $J^{MO2*} \approx 1$, there are some minor instabilities evident from the pressure coefficient plot. For solutions 3 and 4, even though they have a slightly higher $J^{MO1*}$ as compared to solutions 1 and 2, these solutions are more stable and match the pressure coefficient quite closely in the separation and transition region. It is important to note that, while the goal is to improve the boundary layer prediction in the trailing edge region, we do not want to create numerically unstable solutions nor disturb the predictions of properties that are already quite close to the DNS result. Thus, solutions 3 and 4 are quite good solutions to improve the wall shear stress prediction. 
 \begin{figure*}[htb]
\centering
\includegraphics[scale=0.87]{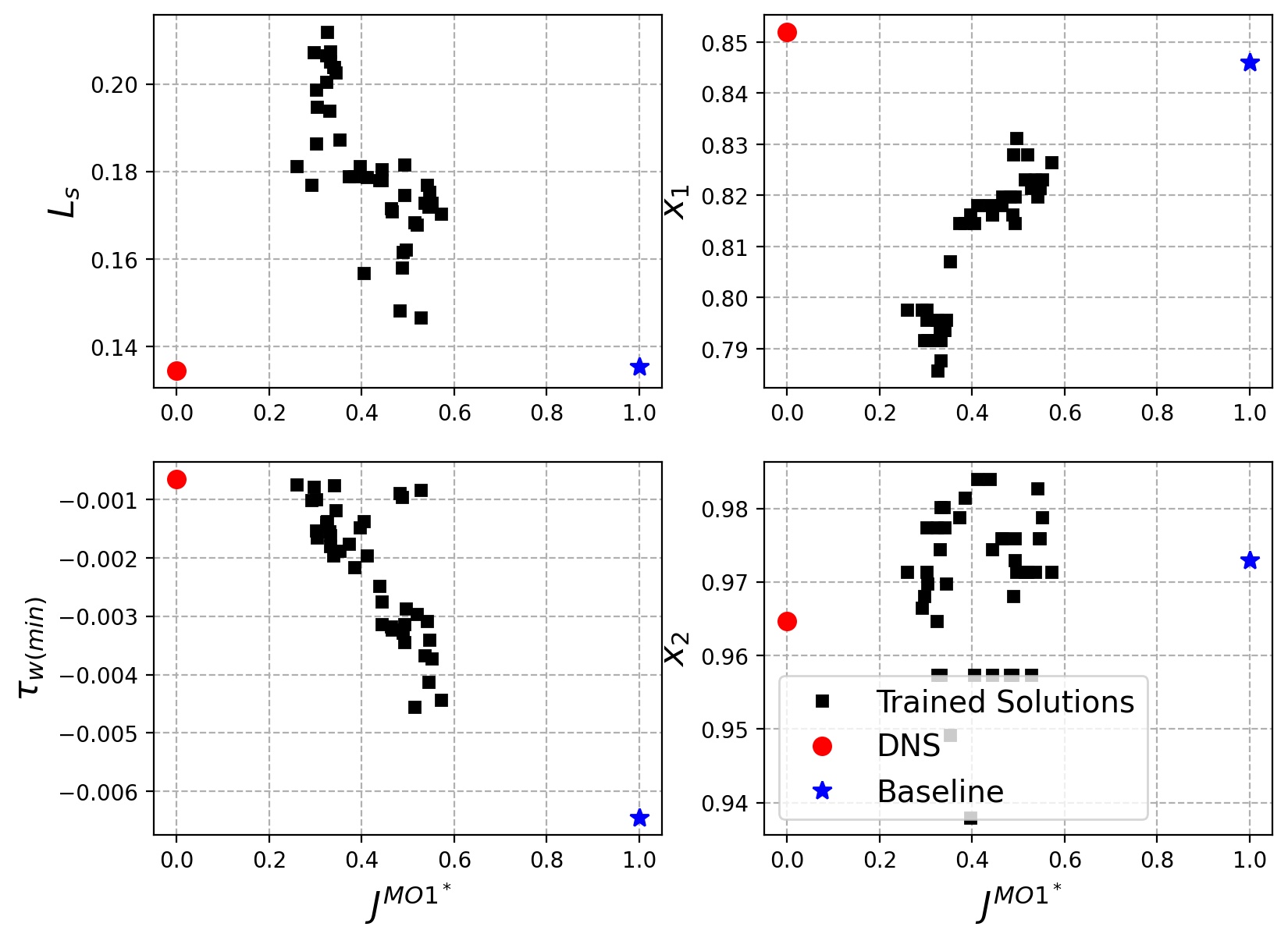}
\caption{The variation in (a) separation bubble length (b) location of separation onset (c) minimum $\tau_w$ in the TE  region (d) Location of transition with respect to the normalised wall shear stress cost function ($J^{MO1*}$).}
\label{wssF}
\end{figure*}

\begin{figure*}[h!]
\centering
\includegraphics[scale=0.87]{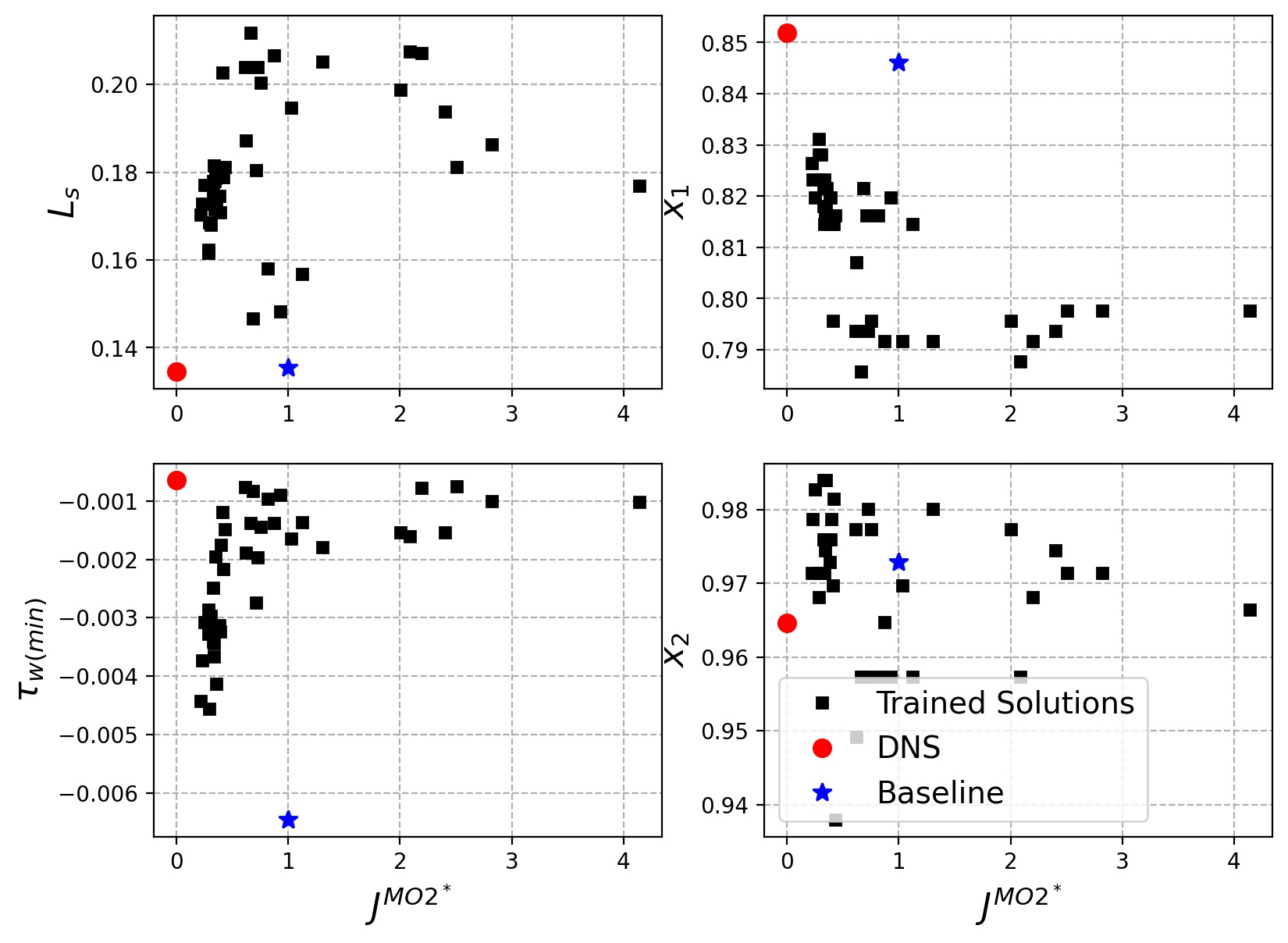}
\caption{The variation in (a) separation bubble length (b) location of separation onset (c) minimum $\tau_w$ in the TE region (d) Location of transition with respect to the normalised pressure coefficient cost function ($J^{M02*}$).}
\label{cpF}
\end{figure*}
Figures 6 and 7 show the variation in the separation bubble length ($L_s$), location of the separation onset ($x_1$), the minimum wall shear-stress, and location of transition onset ($x_2$), with respect to the normalised cost function $J^{M01*}$ and $J^{M02*}$, respectively. In order to locate the exact solution from Fig. 3, one has to see the same sub-plot in Figs. 6 and 7 simultaneously.
From both these plots, it is evident that the new models slightly over-predict the length of the separation bubble as compared to the baseline solution. The trained models also trigger separation slightly earlier than the baseline and DNS cases; hence the longer separation bubble length. However, all the trained values increase the minimum wall shear stress (i.e. reduce the amplitude of the turbulence ignition peak) and bring it much closer to the DNS result from the baseline. A large number of solutions also improve the transition onset location. These last two parameters are all the more crucial for accurately predicting the flow coming off the blade and thus improving the wake mixing prediction. Most of the trained models also improve the predictions (not shown here) in $x_3$ and drastically bring down the peak in the wall shear stress that is present in the baseline solution at $x/C_{ax} = 0.996$. Overall, we can see in terms of the cost functions and parameters outlined in Figs. 6 and 7 the trained solutions have a positive impact on improving the separated flow prediction.  

The model formulation for Solution 3 is:
\begin{equation}
\begin{aligned}
& f_{1a} = \Pi_1 + \Pi_5 - \Pi_6 + \Pi_7 + 0.2115 \\
& f_{2a} = 0.4312 - 0.1534\Pi_2
\end{aligned} \label{eq:model}
\end{equation}
Solution 4 has quite a similar functional form as well, as it lies quite close to Solution 3 in Fig. 3. 
In the above solution, the LKE production term, denoted as $f_{1a}$, underscores the significance of $\Pi_6$. This alignment is in accordance with the assumption made in the foundational LKE model, which posits that the development of pre-transitional fluctuations should be linked to shear stress/strain coupling, as described in reference \cite{lardeau2006}.
The adjustment of $f_{1a}$ concerning the FSTI is achieved by incorporating non-dimensional groups such as $\Pi_5$ in Solution 3. The  $f_{1a}$ term in Eq. \ref{eq:model} can be written as follows,
 \begin{equation}
      f_{1a} = \frac{1}{\Omega} \Big( \frac{k_l + k}{\nu} - \frac{\sqrt{k}}{y} + \omega \Big) + 0.2115
 \end{equation}
Interestingly enough, GEP derived a functional form for  $f_{1a}$ in which all the $\Pi$ group terms had a dependency on the vorticity magnitude. It excluded only those $\Pi$ groups that did not have an inverse relation with the vorticity term in it. This aligns with the original LKE model formulation, in which the laminar eddy viscosity ($\nu_l$) depended on the shear layer vorticity thickness ($\delta_{\Omega}$). The wall distance ($y$) is also incorporated in the model formulation which brings in a dependency on the boundary layer thickness.  
Interestingly enough, GEP derived a functional form for  $f_{1a}$ in which all the $\Pi$ group terms have an inverse dependency on the vorticity.  This aligns with the original LKE model formulation, in which the laminar eddy viscosity ($\nu_l$) depended on the shear layer vorticity thickness ($\delta_{\Omega}$). The wall distance ($y$) is also incorporated in the model formulation which brings the boundary layer thickness into the mix. The new model formulation is more robust than the original formulation as it incorporates a local formulation for $\nu_l$ based on the LKE, TKE and vorticity.

It is noteworthy that solution 3 does not incorporate $\Pi_4$ into the transition sensor or transfer term, contrary to what is specified in the baseline transition model formulation (Eq. 3c). Instead, only $\Pi_2$ is utilized in the functional expression for $f_{2a}$. GEP finds that a function of the shear layer thickness is better attributed to trigger transition onset (sensor) than the wall distance Reynolds number. It also updates the threshold to about 2.8. The updated equation for $\psi$ can be written as 
\begin{equation}
    \psi = max(2.81 - \Pi_2, 0)
\end{equation}
When $\frac{\Omega y}{U}$ is less than 2.81, the transition will be activated. 
To achieve a smooth increase in wall-shear stress as indicated by DNS after the transition onset at $x/C_{ax} = 0.97$ (refer to Fig. 4), one could have anticipated a significant modification of the transfer term compared to the baseline formulation.
The substantial enhancement in the distribution of wall-shear stress observed in the final 10\% of the blade's axial chord, relative to the baseline solution, can be attributed to the behavior of the transfer term.









\section{Conclusions \& Future Work}\label{sec4}
In this study, an attempt was made to improve the prediction of separated flow transition for the T106A LPT blade by enhancing the functional form of the LKE production and transition onset terms with non-dimensional functional forms ($\Pi$ groups). With the help of the CFD-driven MEMO algorithm, a family of best solutions was generated after running the code for 177 generations. The Pareto analysis revealed that only those solutions that improve both the pressure coefficient and wall shear stress prediction over the baseline case are able to offer stable and improved predictions for the separated flow transition. All the solutions generated have a $J^{MO1*}$ which is 40\%-70\% better than the baseline LKE model value. The models developed significantly improved the prediction of transition onset and minimum wall shear stress or turbulence ignition peak, while slightly under-predicting the separation onset. The former has a greater impact on the wake mixing prediction downstream and is of greater importance to blade designers. While the model formulation produced in this study has significantly improved the transition prediction, there is scope to calibrate this model and perform tests on other operating conditions and geometries. 
 
The CFD-driven GEP can also be extended to handle more intricate geometries, like a realistic turbine passage featuring end walls, and also improve wake mixing prediction. 
Such processes could involve minimizing prediction errors related to the endwall vortex structure and wake to mix simultaneously. The GEP framework has already demonstrated its efficacy in enhancing the modeling of complex 3D flows, including junction flows. Nonetheless, its success in such applications would depend on the availability of appropriate training data and the identification of optimal model input features capable of clearly distinguishing various physical phenomena. Both of these aspects constitute elements of ongoing and future research efforts.



\begin{nomenclature}
$C_{ax}$ & Axial chord \\
$C_p$ & Pressure coefficient \\
$f_{1a},f_{2a} $ & Functions to be trained \\
$J$ & Cost Functions \\
$k$ & Turbulent kinetic energy (TKE) \\
$k_l$ & Laminar kinetic energy (LKE) \\
$L_s$ & Length of separation bubble \\
$l_t$ & Turbulence length scale \\ 
$Ma_{2is}$ & Isentropic exit Mach number \\ 
$P_l$ & LKE Production \\
$P_k$ & TKE production \\ 
$R$ & Transfer term \\
$Re_{2is}$ & Isentropic exit Reynolds number  \\
$R_y$ & Wall distance Reynolds number \\
$S$ & Strain-rate magnitude \\
$x_1$ & Separation onset location \\
$x_2$  & Transition onset location \\
$x_3$ & Separation bubble reattachment location \\
$y$ & Wall distance \\
$\delta_{\Omega}$ &  Shear layer vorticity thickness \\
$\psi$ & Transition onset term \\
$\tau_w$ & Wall shear stress \\
$\Pi$ & Non-dimensional Pi groups \\
$\nu_l$ & Laminar Eddy Viscosity \\
$\epsilon$ & Dissipation \\
$\omega$ & Specific dissipation rate \\
$\Omega$ & Vorticity magnitude \\
\end{nomenclature}

 
\end{document}